\documentclass[prl,showpacs,nobibnotes,floatfix,superscriptaddress]{revtex4}

\usepackage{graphicx,eucal}

\begin{document}

\title{Fusing Imperfect Photonic Cluster States}
\author{Yaakov S. Weinstein}
\affiliation{Quantum Information Science Group, {\sc  Mitre}, 260 Industrial Way West, Eatontown, NJ 07724 USA} 
\email{weinstein@mitre.org}

\begin{abstract}
The ability to construct large photonic cluster states capable of supporting universal quantum 
computation relies on fusing together cluster primitives. These fusion operations are probabilistic 
and the efficiency of the construction process relies on recycling remains of cluster primitives 
that have undergone failed fusion attempts. Here I consider the effects of the inevitible decoherence 
that must arise while storing cluster primitives. First, I explore the case where dephased two-qubit 
cluster states are the basic resource for the construction of all larger cluster 
states, all fusion operations are successful, and no further dephasing occurs during the construction 
process. This allows us to explore how decoherence of the most basic, primitive clusters translate into 
imperfections of the larger cluster states constructed from them. I then assume that decoherence occurs 
before every attempted fusion operation and determine the best way to build a five qubit cluster. This 
requires including the effects of the fusion operation failures. Fidelity is
used as the measure of accuracy for the constructed cluster states. Finally, I include a short discussion 
of photon loss and how it affects the construction of simple photonic clusters.
\bigskip
\end{abstract}

\maketitle

\section{Introduction}

Cluster states are highly entangled states that serve as a resource for measurement-based 
universal quantum computation \cite{BR1,BR2,BR3}. A possible experimental venue for cluster 
states is photonics. Nielsen noted \cite{N} that a photonic cluster state quantum computation 
may be more efficient than the original circuit model quantum computation if certain techniques 
from linear optics quantum computation were used to build the photonic cluster. Browne and Rudolph 
\cite{BR} refined this idea replacing Nielsen's construction method with simpler, also probabilistic, 
`fusion' operations. A number of additional methods for constructing clusters have been suggested 
\cite{DR,CCWD} and small photonic cluster states have been experimentally implemented \cite{Zeil,K,Pan}. 

Given that cluster states can, in principle, be built using probabilistic gates, an active area of research 
has been determining the optimal strategy for utilizing cluster resources to build the largest possible 
cluster state \cite{GKE,KGE,KRE,RB,KGE2}. These studies assume the existence of a collection of primitive 
cluster state resources which are fused together to build larger cluster states. A number of strategies 
were suggested and optimal strategies were identified depending on the success probability of the fusion 
gates. All of these studies explicitly or implicitly assume a `pool' or storage unit in which the primitive 
clusters are stored until needed for fusion. This storage unit is assumed to be noiseless: it causes no 
losses or decoherence to any of the stored clusters. 

In this work I relax the assumption that the storage unit is noiseless. Instead, I posit that the storage 
unit causes dephasing (possibly due to birefringence) to the stored clusters. As a first step, I assume 
all two qubit clusters undergo dephasing and then calculate the fidelity of cluster chains of arbitrary 
length built from these basic decohered primitives when fusion gates are always successful. The main part
of the paper will detail the construction of a five qubit cluster assuming that the cluster primitives are
stored in the noisy storage units until needed for a fusion operation and including the possibility 
of fusion failure. Whether the fusion succeeds or fails the resulting cluster(s) are sent back into 
the storage unit until such time as they are again needed for fusion. The result of this study is to 
demonstrate that recycling cluster primitives when a fusion operation fails leads to cluster states 
with significantly lower fidelity than cluster states constructed from only fresh clusters. This is due 
to the increased number of times such primitives must be stored in the noisy storage unit. Taking this 
into account may require revisiting what are optimal cluster construction strategies. 

I choose to explore the construction of a five qubit cluster as it allows us to 
compare two simple construction methods, always fusing two-qubit clusters onto the longest chain and 
fusing the two longest clusters. This is a first step into accounting for the decoherence inherent in 
the storage unit in the cluster construction strategy. The work reported here also complements previous 
work in which it was shown how to utilize the freedom in performing one qubit rotations, a relatively 
easy task in photonic quantum computation, to arrange that the cluster state be stored in a way most 
robust against decoherence \cite{myJMO}. 

To construct clusters I utilize the Type I fusion operations of Browne and Rudolph \cite{BR} 
which is successful 50\% of the time. While initially thought to be applicable only when 
building cluster chains, it has been subsequently shown that Type I fusion alone allows for the 
construction of two-dimensional clusters as well \cite{us}. When the fusion operation is applied 
successfully on the edge qubits of cluster chains of length $m$ and $n$ the output will be a chain 
of length $m+n-1$. Unsuccessful application of the fusion operation will result in the length of 
each chain being reduced by 1.


I first describe the dephasing environment which affects our cluster states during storage. 
I assume no interactions between the qubits in the storage unit. The only dynamics 
in the storage unit is dephasing which is fully described by the Kraus operators
\begin{equation}
K_1 = \left(
\begin{array}{cc}
1 & 0 \\
0 & \sqrt{1-p} \\
\end{array}
\right); \;\;\;\;
K_2 = \left(
\begin{array}{cc}
0 & 0 \\
0 & \sqrt{p} \\
\end{array}
\right),
\end{equation} 
where we have defined the dephasing strength $p$. When all $q$ 
qubits undergo dephasing we have $2^q$ Kraus operators each of the form 
$A_l = (K_i\otimes ... \otimes K_{\ell})$ where 
$l = 1,2,...,2^q$ and $i,...,\ell = 1,2$. 
All of the below calculations are done with respect to $p$, where the exact behavior of 
$p$ as a function of time is left ambiguous so as to accomodate various possible 
dephasing behaviors. As an example, we may assume $p = 1-e^{-\kappa\tau}$ where $\tau$ 
is time and $\kappa$ is the decay constant. In this case, off diagonal terms of the density 
matrix decay as a power of $e^{-\kappa t}$ and thus go to zero (i.~e.~$p\rightarrow 1$) only 
at infinite times.

\section{Construction with Dephased Two-Qubit Clusters}
We start with an arbitrary number of two-qubit clusters which are the most 
basic cluster resource. They are all assumed to be in the state 
$|\psi_2\rangle = \frac{1}{2}(|00\rangle + |01\rangle + |10\rangle - |11\rangle)$. 
The clusters are placed in the noisy storage unit causing them to decohere into the 
state, $\rho_2(p)$, where, in general the dephasing strength $p$ will be dependent on
the amount of time the cluster remains in storage. The dephasing lowers the fidelity,
reduces the purity, and degrades the entanglement of the initially pure two-qubit cluster. 
The fidelity of $\rho_2(p)$ is given by 
$\langle\psi_2|\rho_2(p)|\psi_2\rangle = \frac{1}{4}(2+2\sqrt{1-p}-p)$. The purity of 
$\rho_2$ reduced to $\frac{1}{4}(p-2)^2$ and the entanglement
of the state as quantified by the negativity $N$, defined as the most negative 
eigenvalue of the parital transpose of the density matrix \cite{neg}, behaves as 
$N(p) = \frac{1}{4}(-2\sqrt{1-p}+p)$.

Two two-qubit clusters, both dephased by an amount $p$, are now fused via a Type I fusion 
operation, assumed to be implemented without error. If successful this will result in a three
qubit cluster. However, the dephasing of the resources two qubit clusters causes the fidelity 
of this resulting three-qubit cluster state, $\rho_3(p)$, to be suboptimal, given by: 
$\frac{1}{8}(2+2\sqrt{1-p}-p)(p-2)$. Similarly, the entanglement $N$, of $\rho_3(p)$ is not maximal
and, in fact, may disappear altogether if the storage time is sufficiently long (though finite)
If the negativity is determined with the partial transpose taken with respect to qubits 1 or 3 
$N$ is given by $\frac{1}{8}(-2(1+(1-d)^{3/2})+d(4-d))$, which goes to zero at $p \simeq .7044$. When 
the partial transpose is taken with respect to the second qubit $N$ is given by 
$\frac{1}{8}(-2\sqrt{1-d)(d-2)^2}+2d-d^2)$, which goes to zero at $p = 2(\sqrt{2}-1) \simeq .8284$. 

\begin{figure}
\begin{center}
\includegraphics[height=4cm]{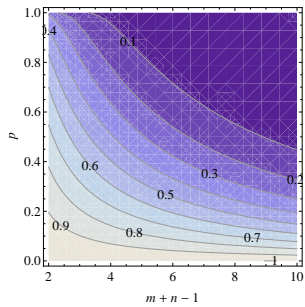}
\caption{\label{fid2}
Fidelity of fused cluster chains using as a resource two-qubit clusters that have 
undergone dephasing of strength $p$. The length of the cluster chain after a successful fusion of
clusters with lengths $m$ and $n$ is $q = m+n-1$.}
\end{center}
\end{figure}

One method of building up to arbitrary length clusters is to continue to fuse two-qubit clusters
onto the one large chain. Thus a five qubit would be fused from a four-qubit cluster, $\rho_4(p)$, 
and $\rho_2(p)$. Another construction method is to fuse together larger cluster states, 
of length $m$ and $n$. Following this method the five-qubit cluster would be constructed from 
two copies of $\rho_3(p)$. Assuming initial resources, $\rho_2(p)$, no further decoherence, and all 
successful fusion operations, the resulting state is the same for both construction methods. 
We can explicitly calculate the fidelity of a $q$-qubit cluster constructed based on the above
assumptions as a function of $p$: 
\begin{equation}
F(q,p) = |\frac{1}{2^q}(2+2\sqrt{1-p}-p)(p-2)^{q-2}|.
\label{primfid}
\end{equation}
This is plotted in Fig.~\ref{fid2}. 

\begin{figure}
\begin{center}
\includegraphics[height=4cm]{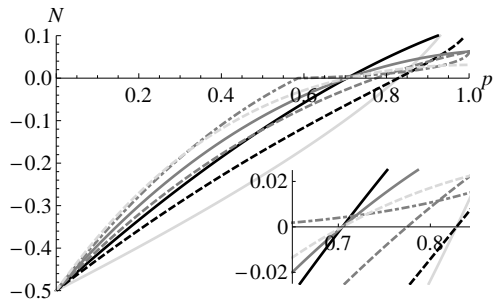}
\caption{\label{ent2}
Negativity of different size cluster states constructed
from basic dephased two-qubit cluster primitives, the two-qubit cluster (light, solid), the constructed 
three-qubit cluster with the partial trace on qubit 1 or 3 (black, solid), on qubit 2 (black, dashed), 
the constructed four qubit cluster with partial trace on qubit 1 (gray, solid), on qubit 2 (gray, dashed), 
on qubits 1 and 2 (gray, chain), constructed five-qubit cluster with partial trace on qubit 1 (light, 
dashed). The inset highlights the decoherence strengths $p = .7044, .8284$ where a number of the 
entanglement measures go to zero despite having different general behavior.}
\end{center}
\end{figure}

Figure \ref{ent2} plots a non-exhaustive group of entanglement measures 
on the constructed cluster states. A number of these measures go to zero (exhibit entanglement 
sudden death \cite{esd}) at the decoherence strengths identified above $p = .7044, .8284$.

\section{Constructing a Five-Qubit Cluster}
We now analyze a more realistic example which assumes the following: (1) the cluster state 
is stored in a noisy storage unit before the application of any fusion operation and thus 
undergoes dephasing of strength $p_t$, (2) the fusion operation may fail, (3) if the fusion operation 
is successful, it works perfectly with no decoherence or errors. Based on this model we attempt to 
determine the most accurate way of constructing a five-qubit cluster state based on the two construction
methods mentioned above: (1) fuse two three-qubit clusters, each of which is built by fusing 
two two-qubit clusters, or (2) fuse a four-qubit cluster and a two-qubit cluster the former of which 
is built by fusing a two-qubit cluster and a three-qubit cluster. As fusion only works 50\% of the time, 
these two construction methods may have different costs in terms of the number of basic two qubit clusters 
used \cite{BR} (as above we assume that two-qubit clusters are readily available). I demonstrate 
that there is also be a difference in the fidelity of the constructed cluster states. 

Both construction methods require (a minimum of) three successful fusion attempts. Let us first assume 
that each of the three fusions are successfully implemented on the first try, which will occur $1/8$th of time. 
In the first construction method, four two-qubit clusters undergo dephasing of strength $p_1$ before being 
successfully fused into two three-qubit clusters. Both three-qubit clusters then undergo dephasing of 
strength $p_2$ before being successfully fused into a five-qubit cluster. 
In the second construction method the single three-qubit cluster undergoes desphasing with strength 
$p_2$ before being fused with another two-qubit cluster (which has undergone dephasing $p_1$). The 
four-qubit cluster then undergoes dephasing $p_3$ before fusion with a final two-qubit cluster (with 
dephasing $p_1$). The fidelities of the constructed states utilizing these two construction methods 
are given explicitly in the Appendix and are plotted in Fig.~\ref{construct} for the cases 
$p_1 = p_2 = p_3 = p$ and $p_1 = 0, p_2 = p_3 = p$ 
(one may argue that this latter scenario is more realistic since two-qubit 
clusters are always available without needing to be stored). In both cases the states 
produced by the two construction methods have practically equal fidelity. 

\begin{figure}
\begin{center}
\includegraphics[width=4cm]{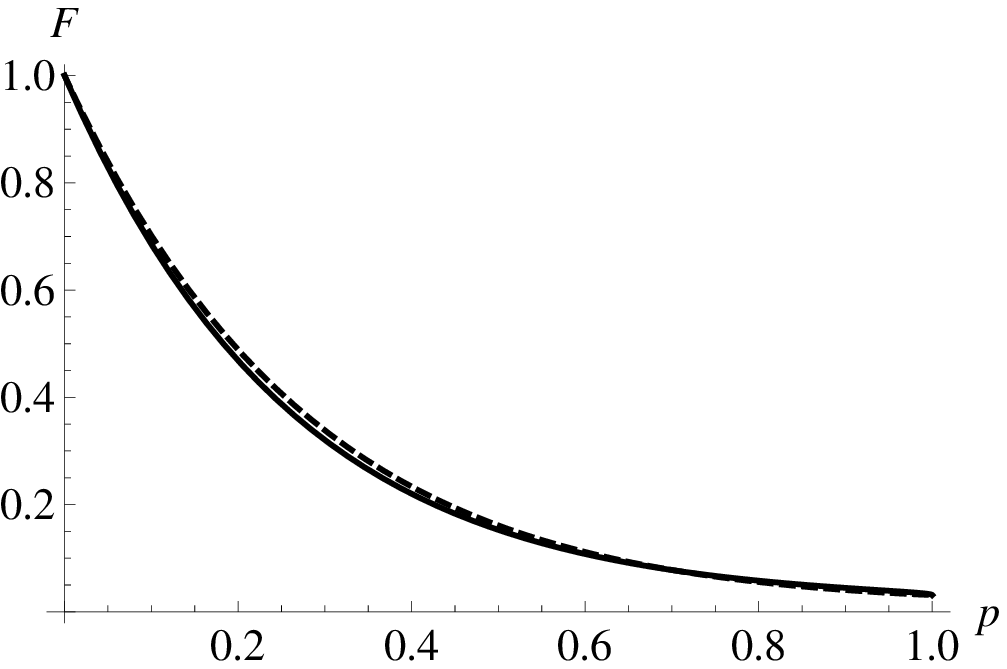}
\includegraphics[width=4cm]{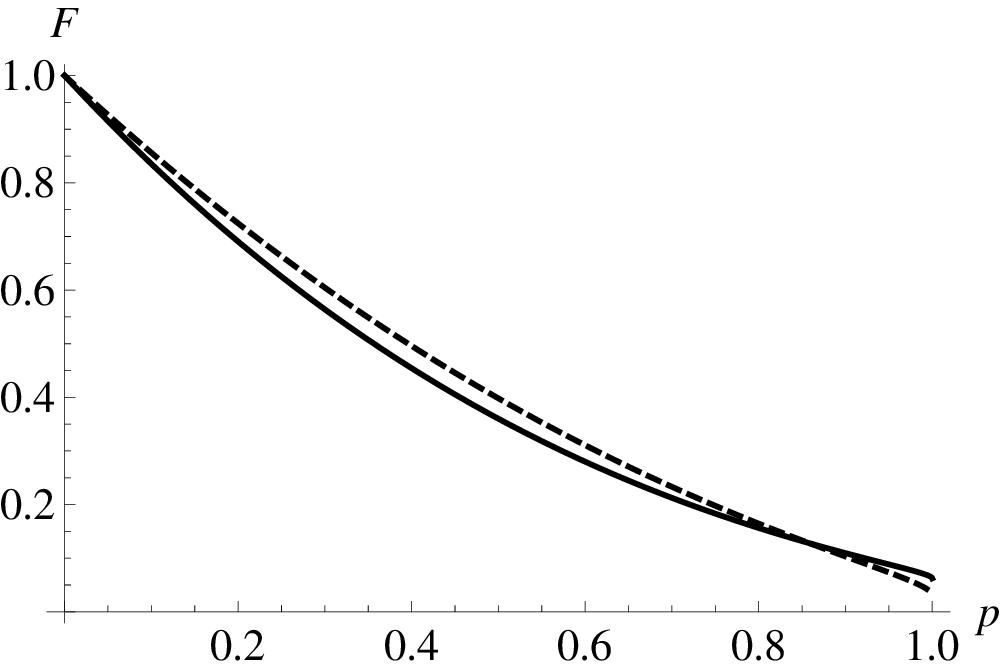}
\caption{\label{construct}
Fidelity of five-qubit cluster states constructed via the two construction methods
(dashed lines for first method, solid lines for second method). Here I assume that the cluster states 
have undergone dephasing before each fusion operation as explained in the text, and that all fusion 
operations have been implemented successfully. Left: all dephasing strengths are equal to $p$. Right:
$p_1 = 0$ and $p_2 = p_3 = p$. }
\end{center}
\end{figure}

In reality, fusion operations do not always work. When a fusion operation fails one could simply discard 
the remaining part of the clusters and start again. In this case success would only be achieved when there
are no fusion failures whatsoever. However, discarding what remains after an unsuccessful fusion would 
greatly decrease the efficiency  of the cluster construction process, as the probability of achieving $f$ 
successful fusion operations without a single failure is $1/2^f$. To attain maximum efficiency requires 
recycling leftover cluster states after a failed fusion attempt. For example, if an attempted fusion 
between two three-qubit clusters has failed the two leftover two-qubit cluster states could be fused 
together into a new three-qubit cluster. The weakness of recycling is that the recycled clusters have 
been stored more times, and thus have undergone more decoherence, than fresh clusters. 

\begin{figure}
\begin{center}
\includegraphics[width=4cm]{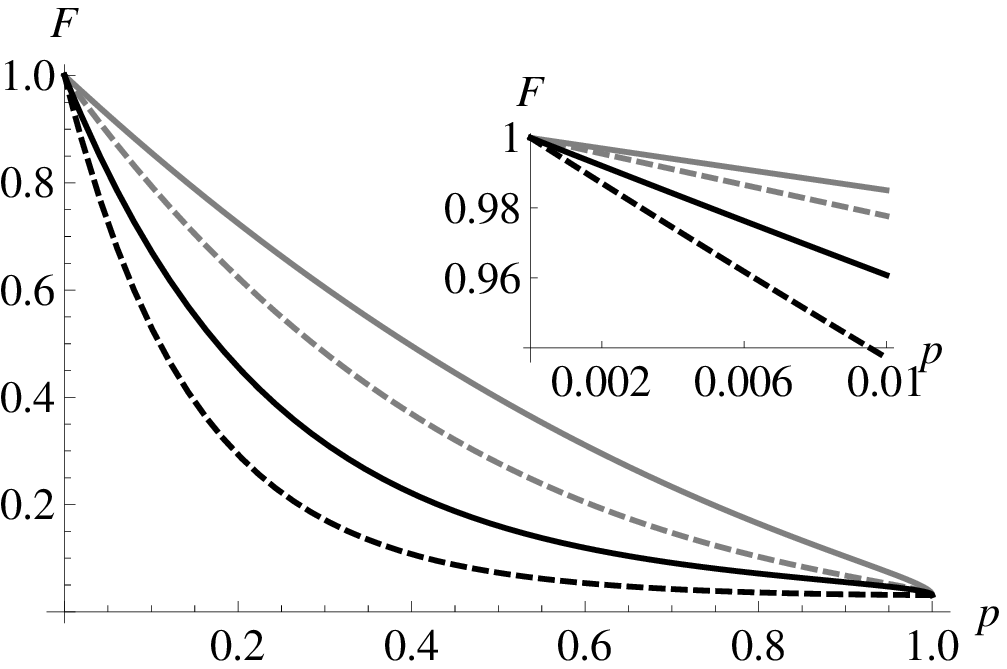}
\includegraphics[width=4cm]{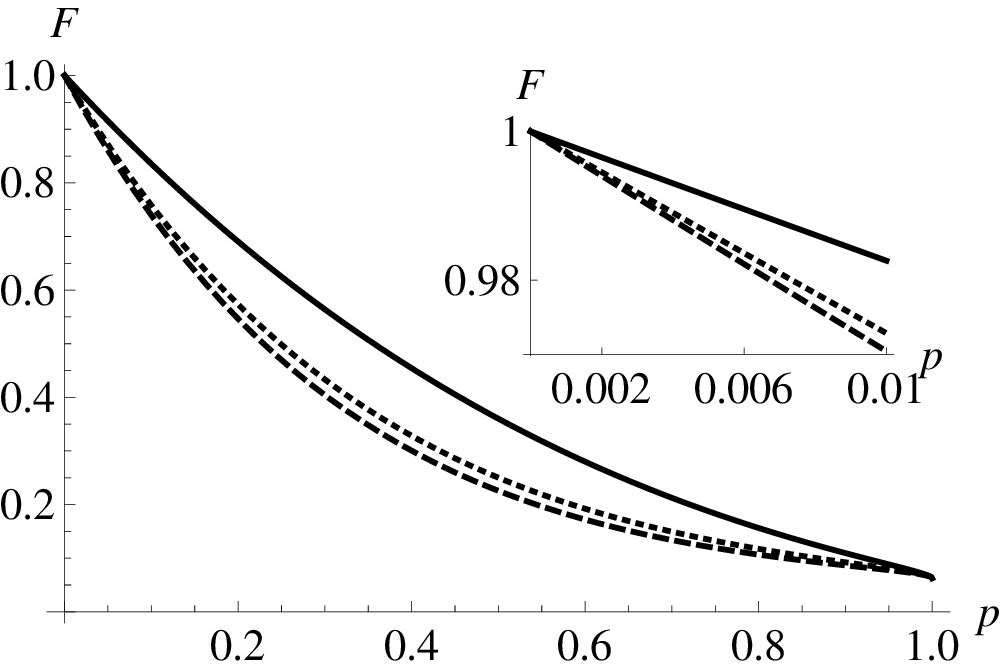}
\caption{\label{recycle}
Left: Fidelity of five-qubit cluster states constructed by fusing together two three-qubit clusters in different
scenarios. The gray lines are cases where the fusion between the two three-qubit clusters succeeds: both 
three-qubit clusters constructed simultaneously (solid), and the three-qubit clusters not constructed 
simultaneously and thus one is subject to dephasing of strength $p_{wait}$ (dashed). The two black lines 
are cases where the fusion between the two three-qubit clusters fails and the two leftover two-qubit 
clusters are successfully recycled by being fused into a new three-qubit cluster:
recycled cluster fused with fresh three-qubit cluster that has not needed any wait (solid), recycled cluster 
fused with another recycled three-qubit cluster (dashed). The dephasing strengths are $p_1 = 0$ and 
$p_2 = p_{wait} = p_3 = p_4 = p$. Inset: fidelity at low values of dephasing.
Right: Fidelity of five-qubit cluster states constructed by fusing together a four-qubit cluster with a two-qubit 
cluster in different scenarios. The solid line is the case where all fusion operations succeed. 
The dotted line is when the fusion between the two- and three-qubit clusters fail and all other fusion 
operations (including that with the recycled two-qubit cluster) succeed. The dashed line is when the 
fusion between the two- and four-qubit clusters fail and all others (including those with the recycled 
three-qubit cluster) succeed. The dephasing strengths are $p_1 = 0$ and $p_2 = p_3 = p_4 = p_5 = p$. 
Inset: fidelity at low values of dephasing.
}
\end{center}
\end{figure}

With this in mind we turn back to our attempt to build a five-qubit cluster. Recall that the first step
in this process must be fusing two two-qubit clusters to form a three-qubit cluster. Were this fusion 
operation to fail what remains would simply be single photons which must be discarded. Thus, an initial 
construction of the three-qubit cluster must be done from two fresh two-qubit clusters and is done with the 
fidelity given in Eq.~\ref{primfid}. The first construction method requires two three-qubit clusters but 
it may not be that both of these clusters are successfully constructed at the same time. If not, one of the 
clusters will have to be stored until the second is constructed. During storage this cluster undergoes
dephasing of strength $p_{wait}$. 
If the wait is too long, such that the fidelity of this three-qubit cluster becomes too low, it may 
not be worth keeping the stored cluster. The two clusters are then fused (after some additional 
storage with dephasing $p_2$ since we assume storage before any fusion operation) to get the 
five-qubit cluster. The fidelity of this cluster as a function of $p_{wait}$ and $p_2$ is shown 
in Fig.~\ref{recycle}.

If the fusion of the two three-qubit clusters fails one could attempt to recycle the leftover pair
of two-qubit clusters by fusing them together to build a new three-qubit cluster. Should this fail
one would have to start all over again but, were it to succeed, one would have a three-qubit cluster 
without consuming any more two-qubit cluster resources. The recycled two-qubit clusters 
undergo another dephasing $p_3$ before their attempted fusion. This `made-from-recyclables' 
three-qubit cluster can then be fused with another three qubit cluster, which may be fresh or recycled, 
after another dephasing $p_4$, to construct the desired five-qubit cluster. 

Figure \ref{recycle} shows the fidelity of the different scenarios we have identified for successful five-qubit 
cluster construction using the first method: successful fusion between the two three-qubit clusters both of 
which are constructed simultaneously, successful fusion between two three-qubit clusters not constructed 
simultaneously (adding dephasing $p_{wait}$), fusion failure with resultant clusters recycled into a 
three-qubit cluster which is then fused with a fresh three-qubit cluster (with dephasing $p_1$, $p_{wait}$, 
$p_2$, $p_3$, and $p_4$), and fusion failure with resultant clusters recycled into a three-qubit cluster 
then fused with another made from recyclable three-qubit cluster. The figure demonstrates that a failure 
in attempting to build one of the three-qubit cluster, thus requiring storage for the successfully fused 
three-qubit cluster, is much less costly, in terms of fidelity, than a failure of the fusion between two 
three-qubit clusters. Explicit expressions for the fidelities (for the case $p_1 = 0$) are given in 
Appendix B. 

We now explore the failure of fusion operations in the second construction method, looking only at 
cases of only one failure. That failure may occur during the second fusion attempt, when fusing the 
three-qubit cluster with a two-qubit cluster, or the third fusion attempt, when fusing the four-qubit 
cluster with a two-qubit cluster. The recycled two- or three-qubit cluster is fused (assumed successfully) 
with additional two-qubit clusters until the five qubit cluster is constructed, requiring a total of five 
fusion attempts. Before each of these five fusion attempts the main cluster undergoes dephasing of strengths
$p_1$,...,$p_5$, while the two-qubit clusters are always assumed to have undergone dephasing $p_1$. 
The fidelity of the cluster states constructed from the recyclables are given in Appendix B and 
plotted in Fig.~\ref{recycle}.
As can be seen, when the failure occurs affects the fidelity of the state very little, though the 
occurence of a failure significantly lowers the fidelity. Additional fusion 
failures can continue to provide recyclable cluster material but the cluster states constructed from 
these recyclables will produce clusters of even lower fidelity.

When deciding which of the above construction methods to use one must take into account the possibility 
of fusion failure. As above, if there are no fusion failures the choice of construction method does not 
seem to matter. Assuming some failure, the effect of storing a three-qubit cluster to wait for another
three-qubit cluster is least damaging to the fidelity of the desired five-qubit cluster. However, a 
failure in the fusion of two three-qubit clusters in the first construction method 
is significantly more harmful than a failure in the fusion the two- and three-qubit clusters or the 
two- and four-qubit clusters of the second construction method. 

All of the above has assumed that the fusion operations, when successful, work perfectly.
In fact, mode mismatch will likely cause errors in the output of even successful fusion operations.
This additional source of error will further penalize construction methods requiring many fusion 
operations and make it less worthwhile to recycle cluster state material as this also requires the 
same states to undergo more fusion operations. Because mode mismatch can modeled by correlated 
Pauli errors \cite{Rohde}, its effect is similar to, and can be determined in a straightforward manner 
from the work reported here.

The above analysis allows us to address the following question: when is it worth 
recycling cluster state material and when is it better to discard everything and start over? 
The answer to this question will depend on the amount of decoherence that the 
clusters experience at every interval and the tolerable fidelity of the desired cluster states. 
What I have shown is simply that using fresh clusters may result in significantly more accurate 
final cluster states than using recycled ones. 

\section{A Word About Photon Loss}
An additional important error mechanism when constructing photonic cluster states is the loss of 
photons. While a complete discussion of photon loss is not within the purview of this work a short discussion is 
included here for the purposes of completeness. A photon can be lost at any step of the cluster state construction
process: at the source, while the cluster is stored, or during the implementation of a fusion 
operation (for exmaple, due to reflection off the surface of a beam-splitter or due to imperfect photon detectors). 

Photon loss can only be detected upon an attempted fusion operation, as this is the only step in the 
construction process that requires measurement. Thus, a loss occuring while the cluster state is in 
the storage unit or at the source to a photon that will never take part in a fusion operation will 
not be detected until the cluster state is used for a computation. If, however, a photon that is to be used 
in a fusion operation is lost, the effects of that loss will depend on the outcome of the attempted fusion 
operation. 

Upon attempted fusion between an actual photon at the edge of a cluster chain and a photon tought to be at the
edge of a separate cluster chain but is actually lost (whether lost at the source, in the storage unit, 
or at the beam-splitter utilized in the fusion operation), at least 50\% of the time (depending on the 
efficiency of the detector) no photon will be detected and the fusion will be presumed to have failed. 
In this case no permanent damage will occur to final cluster state as the lost photon will indeed 
be assumed lost due to fusion failure. If, however, the `non-lost' photon is detected, the fusion operation 
will be deemed successful when in fact there is no photon connecting what were (and actually still are) two 
separate cluster states. If no further fusion operation is attempted this loss will not be detected until 
the cluster state is used for a computation. A similar outcome would occur if two photons taking part in a 
fusion operation enter the detector but only one is detected. It will be assumed that the fusion operation 
succeeded when, in fact, it failed. 

The other possible detector loss scenario is when two actual photons enter the fusion operator and 
one of them enters the detector but is not detected. In this case, it will be assumed that 
the fusion operator failed when, in fact, it was successful. What is thought to be two separate clusters 
of length $m-1$ and $n-1$ is really one cluster of length $(m-1)+1+(n-1) = m+n-1$. Presumably, the fusion 
operation would be tried again using the photons thought to be at the end of each chain: the photons at 
the $m-1$st and $m+1$st places in the actual chain. If this second attempt fails those two photons will be lost along 
with the extra ($m$th) photon that was thought to have been lost in the first fusion attempt. If the second fusion 
attempt is now successful there will be a chain of length $(m-2)+1+(n-2) = m+n-3$. However, the $m-1$st place on
the chain will be comprised of two photons each attached to the photons in places $m-2$ and $m$ and attached to 
each other. How this may affect computational implementations using this cluster will be the subject of future
work. 

In the above few paragraphs I have described the details of how photon loss at the different 
stages of photonic cluster state construction may affect the construction of photonic cluster 
states. Protecting against photon loss is of primary importance to studies of fault tolerance 
using photonic cluster states \cite{H}. Here I have outlined how such loss affects the cluster 
construction on its most basic level, the simplest cluster states. Further investigation is 
necessary to fully develop optimal photonic cluster state construction schemes in the presence 
of dephasing, loss, and other error mechanisms. 

\section{Acknowledgements}
It is a pleasure to acknowledge useful conversations with S. Pappas and support from the MITRE Innovation 
Program under MIP grant \#20MSR053.

\appendix
\section{Cluster Fidelities with no Fusion Failures}
In this appendix we give the explicit expressions for the fidelily of constructed five-qubit 
clusters when there are no fusion failures as a function of the various dephasing strengths. 
For two special cases these fidelities are plotted in Fig.~\ref{construct}. 
In the first construction method the five-qubit cluster is built by fusing two three-qubit
clusters and the fidelity is given by:
\begin{eqnarray}
F_{33} &=& \frac{1}{32}(1+2\tilde{p}_1^{\frac{7}{2}}\tilde{p}_2^{\frac{5}{2}}
+\tilde{p}_1^{4}\tilde{p}_2^{3}+2(\tilde{p}_1\tilde{p}_2)^{\frac{1}{2}}
+2\tilde{p}_1^{\frac{5}{2}}(\tilde{p_2}^{\frac{3}{2}}+\tilde{p}_2^{2})
+2\tilde{p}_1^{3}(\tilde{p}_2^{2}+\tilde{p}_2^{\frac{5}{2}}) \nonumber\\
	 &+& 2\tilde{p}_1(\tilde{p}_2^{\frac{1}{2}}+\tilde{p}_2)(1+(\tilde{p}_1\tilde{p}_2)^{\frac{1}{2}})
+\tilde{p}_1^2\tilde{p}_2(1+\tilde{p}_2+2\tilde{p}_1^{\frac{1}{2}}\tilde{p}_2)
+2\tilde{p}_1^{\frac{3}{2}}(\tilde{p}_2+2\tilde{p}_1^{\frac{1}{2}}\tilde{p}_2^{\frac{3}{2}}),
\end{eqnarray}
where $\tilde{p}_j = 1-p_j$.
In the second construction method the five-qubit cluster is built by fusing a four-qubit cluster
and a two-qubit cluster and the fidelity is given by:
\begin{eqnarray}
F_{24} &=& \frac{1}{32}(1+\tilde{p}_1^{\frac{1}{2}})(1+(\tilde{p}_1\tilde{p}_2\tilde{p}_3)^{\frac{1}{2}}
+\tilde{p}_1^{\frac{1}{2}}(\tilde{p}_1^{\frac{3}{2}}(\tilde{p}_2^{\frac{1}{2}}+\tilde{p}_2)\tilde{p}_3
+\tilde{p}_1(2\tilde{p}_2+(\tilde{p}_1\tilde{p}_2)^{\frac{1}{2}})\tilde{p}_3 \nonumber\\
&+&\tilde{p}_1^3\tilde{p}_2^{\frac{3}{2}}\tilde{p}_3^{2}
+(\tilde{p}_1\tilde{p}_3)^{\frac{1}{2}}+(\tilde{p}_1\tilde{p}_2\tilde{p}_3)^{\frac{1}{2}}+\tilde{p}_1^{\frac{1}{2}}
((\tilde{p}_1\tilde{p}_2)^{\frac{1}{2}}\tilde{p}_3+(\tilde{p}_2\tilde{p}_3)^{\frac{1}{2}}) \nonumber\\
&+&\tilde{p}_1^2\tilde{p}_2\tilde{p}_3(2\tilde{p}_3^{\frac{1}{2}}+(\tilde{p}_1\tilde{p}_3)^{\frac{1}{2}}+(\tilde{p}_2\tilde{p}_3)^{\frac{1}{2}}))).
\end{eqnarray}

\section{Cluster Fidelities with Fusion Failures}
In this appendix we give the explicit expressions for the fidelily of constructed five-qubit 
clusters when there are fusion failures as a function of the various dephasing strengths with
$p_1 = 0$. For the special case of all dephasing strengths equal these fidelities are plotted 
in Fig.~\ref{construct}. In the first construction method the five-qubit cluster is built by 
fusing two three-qubit and we have three failure possibilites: when the three-qubit clusters 
are not constructed simultaneously, when one of the three-qubit clusters is made from recylced
cluster material and when both three qubit clusters are made from recycled cluster materials. 
The three fidelities are given by:
\begin{equation}
F_{33Wait} = \frac{1}{32}(1+\tilde{p}_2^{\frac{1}{2}})^2(1+\tilde{p}_2\tilde{p}_w+2(\tilde{p}_2\tilde{p}_w)^{\frac{1}{2}}
+\tilde{p}_2\tilde{p}_w^{\frac{1}{2}}+\tilde{p}_2^{\frac{3}{2}}\tilde{p}_w(2+(\tilde{p}_2\tilde{p}_w)^{\frac{1}{2}})),
\end{equation}
\begin{eqnarray}
F_{FailFresh} &=& \frac{1}{32}(1+\tilde{p}_4^{\frac{1}{2}})^2(1+(\tilde{p}_2\tilde{p}_3)^{\frac{1}{2}}\tilde{p}_4
+(\tilde{p}_2\tilde{p}_3)^{\frac{3}{2}}\tilde{p}_4\tilde{p}_w+\tilde{p}_2^2\tilde{p}_3^2\tilde{p}_4^2\tilde{p}_w\nonumber\\
&+& (\tilde{p}_2\tilde{p}_3+(\tilde{p}_2\tilde{p}_3)^{\frac{1}{2}})(\tilde{p}_2\tilde{p}_3\tilde{p}_w)^{\frac{1}{2}}\tilde{p}_4^{\frac{3}{2}}
+\tilde{p}_2\tilde{p}_3(\tilde{p}_4\tilde{p}_w)^{\frac{1}{2}}+(\tilde{p}_2\tilde{p}_3\tilde{p}_4\tilde{p}_w)^{\frac{1}{2}}),
\end{eqnarray}
\begin{eqnarray}
F_{FailFail} &=& \frac{1}{32(\tilde{p}_2\tilde{p}_3)^{\frac{1}{2}}}((\tilde{p}_2\tilde{p}_3)^4(1
+(\tilde{p}_2\tilde{p}_3\tilde{p}_4)^{\frac{1}{2}})\tilde{p}_4^{\frac{5}{2}}\tilde{p}_w^2+\tilde{p}_2^3\tilde{p}_3^3\tilde{p}_4^2\tilde{p}_w(1
+2(\tilde{p}_2\tilde{p}_3\tilde{p}_w)^{\frac{1}{2}}\nonumber\\
&+& (2+(\tilde{p}_2\tilde{p}_3)^{\frac{1}{2}})(\tilde{p}_2\tilde{p}_3\tilde{p}_4\tilde{p}_w)^{\frac{1}{2}})
+\tilde{p}_2^2\tilde{p}_3^2\tilde{p}_4\tilde{p}_w^{\frac{1}{2}}(2+\tilde{p}_4^{\frac{1}{2}}+(\tilde{p}_4\tilde{p}_w)^{\frac{1}{2}}\nonumber\\
&+& 2(\tilde{p}_2\tilde{p}_3\tilde{p}_4\tilde{p}_w)^{\frac{1}{2}})+(\tilde{p}_2\tilde{p}_3)^{\frac{1}{2}}(1+(\tilde{p}_2\tilde{p}_3\tilde{p}_4)^{\frac{1}{2}}
+2\tilde{p}_2\tilde{p}_3(\tilde{p}_4\tilde{p}_w)^{\frac{1}{2}}+(\tilde{p}_2\tilde{p}_3\tilde{p}_4\tilde{p}_w)^{\frac{1}{2}}\nonumber\\
&+&2\tilde{p}_2\tilde{p}_3\tilde{p}_4\tilde{p}_w^{\frac{1}{2}}+(\tilde{p}_2\tilde{p}_3)^{\frac{3}{2}}\tilde{p}_4\tilde{p}_w
+\tilde{p}_2^2\tilde{p}_3^2\tilde{p}_4\tilde{p}_w(1+(\tilde{p}_2\tilde{p}_3)^{\frac{1}{2}}+2(\tilde{p}_2\tilde{p}_3\tilde{p}_4)^{\frac{1}{2}}\nonumber\\
&+& (\tilde{p}_2\tilde{p}_3)^{\frac{1}{2}}\tilde{p}_4(1+(\tilde{p}_2\tilde{p}_3\tilde{p}_w)^{\frac{1}{2}})+\tilde{p}_2\tilde{p}_3\tilde{p}_4^{\frac{1}{2}}(1+\tilde{p}_w^{\frac{1}{2}})
+\tilde{p}_2\tilde{p}_3\tilde{p}_4(1+\tilde{p}_w^{\frac{1}{2}}))).
\end{eqnarray}

In the second construction method the the five-qubit cluster is built by 
fusing a four-qubit and a two qubit cluster. We examine two failure possibilites: when the 
fusion of the three-qubit and two-qubit clusters fail and when the fusion of the four-qubit 
and a two-qubit clusters fail. In each case the five qubit cluster is constructed from once
recycled cluster material. The two fidelities are given by:
\begin{eqnarray}
F_{3Fail} &=& \frac{1}{256}((16\tilde{p}_2\tilde{p}_3+5(\tilde{p}_2\tilde{p}_3)^{\frac{1}{2}})\tilde{p}_4\tilde{p}_5^{\frac{3}{2}}
+8\tilde{p}_2\tilde{p}_3\tilde{p}_4^{\frac{3}{2}}\tilde{p}_5^2+2\tilde{p}_5^{\frac{1}{2}}(8+\tilde{p}_4(\tilde{p}_2\tilde{p}_3\tilde{p}_5)^{\frac{1}{2}})\nonumber\\
&+& 2(8+8(\tilde{p}_4\tilde{p}_5)^{\frac{1}{2}}+16(\tilde{p}_2\tilde{p}_3\tilde{p}_4\tilde{p}_5)^{\frac{1}{2}}
+(\tilde{p}_2\tilde{p}_3\tilde{p}_4)^{\frac{1}{2}}\tilde{p}_5+(\tilde{p}_2\tilde{p}_3)^{\frac{1}{2}}\tilde{p}_4\tilde{p}_5)\nonumber\\
&+& \tilde{p}_5(16\tilde{p}_4^{\frac{1}{2}}+2(15(\tilde{p}_2\tilde{p}_3\tilde{p}_4)^{\frac{1}{2}}+(\tilde{p}_2\tilde{p}_3)^{\frac{1}{2}}\tilde{p}_4
+2(\tilde{p}_2\tilde{p}_3\tilde{p}_5)^{\frac{1}{2}}\tilde{p}_4)+\tilde{p}_4(26(\tilde{p}_2\tilde{p}_3)^{\frac{1}{2}}\nonumber\\
&+& 23(\tilde{p}_2\tilde{p}_3\tilde{p}_5)^{\frac{1}{2}}
+8\tilde{p}_2\tilde{p}_3(2+2(\tilde{p}_4\tilde{p}_5)^{\frac{1}{2}}+\tilde{p}_4^{\frac{1}{2}}\tilde{p}_5))))
\end{eqnarray}
\begin{eqnarray}
F_{4Fail} &=& \frac{1}{128}(8+8\tilde{p}_5^{\frac{1}{2}}+(3\tilde{p}_2^{\frac{1}{2}}+8\tilde{p}_2)\tilde{p}_3\tilde{p}_4\tilde{p}_5^{\frac{3}{2}}
+4\tilde{p}_2(\tilde{p}_3\tilde{p}_4)^{\frac{3}{2}}\tilde{p}_5^2+8(\tilde{p}_3\tilde{p}_4\tilde{p}_5)^{\frac{1}{2}}\nonumber\\
&+& 16(\tilde{p}_2\tilde{p}_3\tilde{p}_4\tilde{p}_5)^{\frac{1}{2}}+(\tilde{p}_2\tilde{p}_3\tilde{p}_4)^{\frac{1}{2}}\tilde{p}_5
+ \tilde{p}_5(8(\tilde{p}_3\tilde{p}_4)^{\frac{1}{2}}+15(\tilde{p}_2\tilde{p}_3\tilde{p}_4)^{\frac{1}{2}}\nonumber\\
&+& \tilde{p}_3\tilde{p}_4(16\tilde{p}_2^{\frac{1}{2}}+13(\tilde{p}_2\tilde{p}_5)^{\frac{1}{2}}+4\tilde{p}_2(2+2(\tilde{p}_3\tilde{p}_4\tilde{p}_5)^{\frac{1}{2}}
+(\tilde{p}_3\tilde{p}_4)^{\frac{1}{2}}\tilde{p}_5)))).
\end{eqnarray}

\end{document}